\documentclass[preprint,12pt]{elsarticle}

\usepackage{url}
\usepackage{upgreek}
\usepackage{graphicx}         
\usepackage{amsmath}
\usepackage{amsfonts}
\usepackage{algorithm}
\usepackage{algpseudocode}
\usepackage{bm}   
\usepackage{amssymb}
\usepackage{upgreek}
\usepackage{dsfont}
\usepackage{enumitem}
\usepackage[dvipsnames]{xcolor}
\usepackage{booktabs}
\usepackage{diagbox}
\usepackage{pifont}

\newcommand{\cmark}{\ding{51}}
\newcommand{\xmark}{\ding{55}}

\newtheorem{prop}{Proposition}
\newtheorem{thm}{Theorem}
\newtheorem{lem}{Lemma}
\newdefinition{defn}{Definition}
\newdefinition{rem}{Remark}
\newdefinition{assum}{Assumption}
\newproof{pf}{Proof}

\newcommand{\eyem}{\mf{I}}

\newcommand{\sgn}[2]{\left\lceil#1\right\rfloor^{#2}}
\newcommand{\diag}[1]{\mbox{diag}(#1)}
\newcommand{\cov}{\normalfont\textsf{\footnotesize cov}}

\newcommand{\mfs}[1]{{\normalfont\textsf{#1}}}
\newcommand{\mf}{\mathbf}

\def\RevOne#1{\textcolor{black}{#1}}
\def\RevTwo#1{\textcolor{black}{#1}}
\def\RevThree#1{\textcolor{black}{#1}}
\def\RevAll#1{\textcolor{black}{#1}}

\begin{document}

\begin{frontmatter}

\title{Perception-latency aware distributed target tracking \tnoteref{t1}}

\tnotetext[t1]{This work was supported via projects PID2021-124137OBI00 and TED2021-130224B-I00 funded by MCIN/AEI/10.13039/501100011033, by ERDF A way of making Europe and by the European Union NextGenerationEU/PRTR, by the Gobierno de Aragón under Project DGA T45-20R, by the Universidad de Zaragoza and Banco Santander, by the Consejo Nacional de Ciencia y Tecnología (CONACYT-Mexico) with grant number 739841.
\textcolor{red}{This is the accepted version of the manuscript: R. Aldana-López, R. Aragüés, and C. Sagüés, ``Perception-latency aware distributed target tracking," Information Fusion, vol. 99, p. 101857, 2023, ISSN 1566-2535. doi: 10.1016/j.inffus.2023.101857.
    \textbf{Please cite the publisher's version}. For the publisher's version and full citation details see:\\\protect\url{https://doi.org/10.1016/j.inffus.2023.101857}.}
}

\author[First]{Rodrigo Aldana-López*} 
\author[First]{Rosario Aragüés} 
\author[First]{Carlos Sagüés} 

\address[First]{Departamento de Informatica e Ingenieria de Sistemas (DIIS) and Instituto de Investigacion en Ingenieria de Aragon (I3A), 
\\
Universidad de Zaragoza, Zaragoza 50018, Spain.\\
(e-mail: rodrigo.aldana.lopez@gmail.com, raragues@unizar.es, csagues@unizar.es)}

\begin{abstract}       
This work is devoted to the problem of distributed target tracking when a team of robots detect the target through a variable perception-latency mechanism. A reference for the robots to track is constructed in terms of a desired formation around the estimation of the target position. However, it is noted that due to the perception-latency, classical estimation techniques have smoothness issues which prevent asymptotic stability for the formation control. We propose a near-optimal smooth-output estimator which circumvents this issue. Moreover, local estimations are fused using novel dynamic consensus techniques. The advantages of the proposal as well as a comparison with a non-smooth optimal alternative are discussed through simulation examples.
\end{abstract}

\end{frontmatter}

\section{Introduction}

Formation control of mobile robots for target tracking is a problem of great interest nowadays with a wide range of applications such as automated surveillance \cite{escort, fusion_trajectory}, aerial filming using drones \cite{film}, intelligent transportation \cite{transport}, among other examples. In the most general setting, the goal is to coordinate a team of robots in order to maintain a formation around a moving target. This can be performed by detecting the target locally at each robot, fuse estimations in a decentralized fashion, and then use the resulting target estimation as a reference to achieve the formation.

Target localization is usually performed by the robots using a Visual Target Detector (VTD). Popular recent examples of VTD algorithms include \cite{redmon2016,beery2020,ren2015,Dai2016,cyber_rnn}, which mostly rely on Deep Learning (DL) models. For these type of algorithms, the term perception-latency is used to denote the computing time employed starting from the sampling instant for the vision sensors to the moment in which the vision algorithms produce an output \cite{aldana2020}. Note that the quality of the perception output can be improved by increasing the image resolution, the number of features, using more complex feature descriptors or using different DL models depending on the vision algorithm. However, these improvements come at the expense of increasing the perception-latency as well. Such perception-latency and detection quality trade-off has been characterized for some methods as reported in \cite{huang2017}. 

In general, resource-constrained robots will trade-off between accuracy and speed for target detection. Some approaches allow to choose the perception latency as requested online for resource economy or to improve the accumulated estimation error. This can be performed either by using a bank of perception methods \cite{aldana2020,luo2019,guan2018}, or by using Anytime Neural Networks (ANN) as in \cite{scheduling,hu2019}. In some cases, the perception latency may be too large when compared with the actual dynamics of the robot. Therefore, state estimation and prediction must be performed locally in order to obtain a target estimate in-between measurements. However, due to the hybrid nature of the problem, if standard techniques such as a Kalman filter are used, the resulting estimation will be discontinuous whenever a new measurement is used for correction. As we show later in this manuscript, if such estimation is used as a reference for the robot to track, the discontinuities in the trajectory may prevent asymptotic stability for the formation control task.

A similar issue occurs in the multi-robot setting. Several strategies have been proposed in the literature in order to fuse local Kalman filter estimations either using a fusion center as in \cite{fusion_filtering1,fusion_filtering2} or in a decentralized fashion as is of interest in this work. In this context, in \cite{static_consensus} local Kalman filter estimations are combined using discrete-time static consensus protocols, ignoring cross-correlations between agents. However, these only achieve exact convergence when an infinite number of iterations for the consensus protocol are performed at each sampling step. In contrast, the works \cite{olfati1,olfati2} use linear dynamic consensus filters \cite{Solmaz2017} which do not have the previously mentioned issue. Despite this, as discussed in \cite{edcho}, linear dynamic consensus protocols cannot exhibit exact convergence with persistently varying references. Other recent fusion strategies have been proposed which improve the fusion quality by using the covariance intersection method \cite{CI1,CI2,CI3}. The issue with all of the previously mentioned fusion methods is that they mainly work in discrete time. This means that synchronous measurements and updates might be assumed, which is incompatible with the perception-latency setting. On the other hand, the discontinuity issue in the estimations prevails in all these methods, since a continuous time-prediction must be computed in-between filter updates as well.

Motivated by this discussion we contribute \RevAll{with a hybrid fusion} framework, which obtains near-optimal smooth estimations for the target, regardless of the perception-latency mechanism. This means that the control design is decoupled from the perception configuration decisions, avoiding any stability issues arising from the discontinuous estimations. Moreover, we leverage the recently developed dynamic consensus protocol \cite{redcho} which, in contrast to other linear dynamic consensus protocols, achieves exact convergence under mild assumptions. The advantages of the proposal as well as a comparison with a non-smooth alternative are discussed through simulation examples.

\subsection{Notation}
Let $\binom{\mu}{\nu}$ denote the binomial coefficient. Let $\text{sign}(x) = 1$ if $x> 0$, $\text{sign}(x)=-1$ if $x<0$ and $\text{sign}(0)=0$. Moreover, if $x\in\mathbb{R}$, let $\lceil x\rfloor^\alpha:= |x|^\alpha\text{sign}(x)$ for $\alpha>0$ and $\lceil x\rfloor^0:={\text{sign}}(x)$. When $\mf{x}=[x_1,\dots,x_n]^T\in\mathbb{R}^n$, then $\sgn{\mf{x}}{\alpha}:=\left[\sgn{x_1}{\alpha},\dots,\sgn{x_n}{\alpha}\right]^T$ for $\alpha\geq 0$ and similarly for $\sgn{\mf{A}}{\alpha}, \mf{A}\in\mathbb{R}^{n\times n}$. Let $\mathbb{E}\{\bullet\}, \cov\{\bullet\}$ denote expectation and covariance operators respectively. For any function $u(t)$, let $u^{(\mu)}(t)$ for $\mu\in\mathbb{N}$ denote its $\mu$-th derivative when it exists.

\section{Problem statement}

Consider a team of $\mfs{N}$ mobile robots located at positions $\mathbf{p}_i(t)\in\mathbb{R}^n, i\in\{1,\dots,\mfs{N}\}$ where $n$ is the dimension of the workspace of the robots. For simplicity, each robot is modeled to be holonomic with $m$-th order integrator dynamics  
\begin{equation}
\label{eq:agents}
\dot{\mf{x}}_i(t) = \mf{A}\mf{x}_i(t) + \mf{B}\mf{u}_i(t),
\end{equation}
where $\mf{x}_i(t) = [(\mathbf{p}_i^{(0)}(t))^\top ,\dots,(\mathbf{p}_i^{(m-1)}(t))^\top ]^\top$, $\mf{u}_i(t)\in\mathbb{R}^n$ is a local control input and $\mf{A},\mf{B}$ are given in \ref{ap:aux_mat}. We assume that each robot is able to measure its own state $\mf{x}_i(t)$ using local sensors. The robots are able to share information between them according to a communication network modeled by an undirected graph $\mathcal{G}$. 
In addition, a target of interest with position $\mathbf{p}(t)\in \mathbb{R}^n$ is assumed to have dynamics with similar integrator dynamics as in \eqref{eq:agents}. Moreover, since the input at the target is unknown, we model it in a stochastic fashion as
\begin{equation}
\label{eq:target}
\text{d} \mf{x}(t) = \mf{A}\mf{x}(t)\text{d}t + \mf{B}\text{d}\mathbf{u}(t)
\end{equation}
where $\mf{p}(t)=\mf{C}\mf{x}(t)$ with $\mf{C}$ defined in \ref{ap:aux_mat} and $\mf{u}(t)\in\mathbb{R}^n$ is a $n$-dimensional Wiener processes with covariance $\cov\{\mf{u}(s),\mf{u}(r)\} = \mf{W}\min(s,r)$ \cite[Page 63]{astrom}. As usual, the process $\mf{u}(t)$ models disturbances, unknown inputs at the target, and non-modeled dynamics. 

\textbf{Perception mechanism}: The robot $i$ uses available sensors  such as vision, range, etc,  to produce \textit{raw measurements} of the environment. For each raw measurement, a \textit{processed
measurement} for the position of the target is obtained through a detection process at, perhaps non-uniform, processing instants $\bm{\tau}_i=\{\tau_{i,k}\}_{k=0}^\infty$, $\tau_{i,0}=0$.  Processed measurements are detections of $\mf{p}(\tau_{i,k}) = \mf{C}\mf{x}(\tau_{i,k})$. According to the current processing and energy budget of robot $i$, a perception latency $\Delta_{i,k}\in[\Delta_{\min}, \Delta_{\max}]$ with $\Delta_{\min}, \Delta_{\max}>0$ is chosen for the detection process at $t=\tau_{i,k}$ and the processed measurement $\mf{z}_i(\tau_{i,k}) = \mf{C}\mf{x}(\tau_{i,k}) + \mf{v}_i(\tau_{i,k})$ is available at $t=\tau_{i,k} + \Delta_{i,k}$. Here, $\mf{v}_i(\tau_{i,k})$ is a Gaussian noise modeling the accuracy of the perception method. In general, $\mf{R}(\Delta_{i,k})=\cov\{\mf{v}_i(\tau_{i,k})\}$ decreases in magnitude as more processing time $\Delta_{i,k}$ is employed.

\begin{rem}
In order to focus in the estimation and information fusion aspects of this work, we consider that the process for which the perception latencies $\{\Delta_{i,k}\}_{k=0}^\infty$ are chosen at robot $i$ is already given. For instance, some ideas from \cite{aldana2020,codesign} can be applied where only $D$ available perception methods can be used, and the perception-latency schedule $\Delta_{i,k}\in\{\Delta^j\}_{j=1}^D$ is chosen to minimize some local performance function. 
\end{rem}

The goal is for the robots to achieve a given formation provided known displacements $\mf{d}_{1},\dots,\mf{d}_{\mfs{N}}\in\mathbb{R}^n$ around the target. However, given that target localization is imperfect, an approximate scheme must be adopted. The strategy is to obtain local estimates $\hat{\mf{x}}_i(t)$ of the target state at each robot given the local perception-latency schedule $\{\Delta_{i,k}\}_{k=0}^\infty$, and then combine them into a single global estimate $\bar{\mf{p}}_{\mfs{G}}(t)$ for the trajectory $\mf{p}(t)$ in a distributed and decentralized fashion. Then, achieve a formation around $\bar{\mf{p}}_{\mfs{G}}(t)$ instead. This is, given a 
reference $\mf{p}_{i}^{\mfs{r}}(t)=\bar{\mf{p}}_\mfs{G}(t)+\mf{d}_i$, reach
\begin{equation}
\label{eq:goal}
\text{\textbf{Goal:\ }}\lim_{t\to\infty}\|\mf{p}_i(t)-\mf{p}_{i}^{\mfs{r}}(t)\|=0, \forall i\in\{1,\dots,\mfs{N}\}
\end{equation}
The reference should be smooth enough for $(\mf{p}_{i}^{\mfs{r}})^{(m)}(t), \forall t\geq 0$ to exist and achieve trajectory tracking at each robot. Nonetheless, this is incompatible with existing optimal estimation techniques taking into account the hybrid nature of the problem. 

\subsection{\RevOne{Solution outline}}
\label{sec:outline}
To solve the problem, the proposal contains the following ingredients:
\begin{itemize}
    \item \textbf{Smooth-output estimation}:  We propose an alternative to classical filtering obtaining a smooth trajectory of the estimate $\hat{\mf{x}}_i(t)$ of ${\mf{x}}(t)$ locally at robot $i$.
    \item \textbf{Estimation fusion}: A distributed and decentralized consensus filter is used to fuse the local information of $\hat{\mf{x}}_i(t)$ into the single estimate $\bar{\mf{p}}_{\mfs{G}}(t)$ of $\mf{p}(t)$. 
    \item \textbf{Formation control:} A local controller $\mf{u}_i(t)$ is designed for the robot $i$ to achieve the formation goal in \eqref{eq:goal} using the estimation framework.
\end{itemize}

\begin{figure}
\centering
\includegraphics[width=0.9\textwidth]{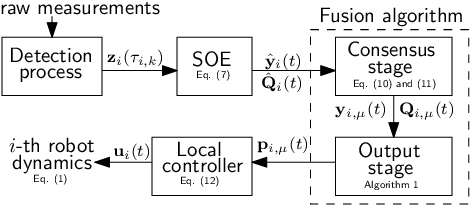}
\caption{\RevOne{High-level of the processing flow for our proposal as described in Section \ref{sec:outline}.}}
\label{fig:outline}
\end{figure}

\RevOne{Figure \ref{fig:outline} presents a high-level outline of the processing flow for our proposal. In particular, we assume that the target detection process is already given, taking raw measurements of the environment and producing detections for the target as $\mf{z}_i(\tau_{i,k})$. The Smooth-Output Estimation (SOE) block is described in detail in Section \ref{sec:smooth_output}, which takes the output of the detection process, and computes smooth information vector and matrix $\hat{\mf{y}}_i(t),\hat{\mf{Q}}
_i(t)$ associated with the estimation $\hat{\mf{x}}_i(t)$ and its covariance $\hat{\mf{P}}_i(t)$ for the state of the target. The fusion algorithm is described in detail in Section \ref{sec:fusion}, and is composed of two modules. First, a consensus stage computes a fused version of the information vector and matrix ${\mf{y}}_{i,0}(t),{\mf{Q}}_{i,0}(t)$ for the target state as well as its derivatives $\{{\mf{y}}_{i,\mu}(t),{\mf{Q}}_{i,\mu}(t)\}_{\mu=1}^m$. Then, the fusion output stage uses them to compute the actual joint estimation $\bar{\mf{p}}_{\mfs{G}}(t)$ for the target position $\mf{p}(t)$ and its first $m$ derivatives. These estimations are stored in the local variables $\{\mf{p}_{i,\mu}(t)\}_{\mu=0}^m$.  Finally, these signals are used to compute a local trajectory tracking control for each robot, which is described in Section \ref{sec:control}.}

\RevOne{The main advantage of the architecture depicted in Figure \ref{fig:outline}, is that the smooth estimation block allows the fusion and control blocks to be designed independently from the detection procedure, providing more versatility than in prior literature. In addition, the smooth output estimation and fusion greatly reduces the estimation and tracking errors as shown in the experiments provided in Section \ref{sec:simulations}.}

\section{Smooth-output estimation}
\label{sec:smooth_output}
In this section, we focus on the local target estimation at each robot. To simplify the presentation, we drop the index $i$ when there is no ambiguity given that all variables are assumed to be local. First, given a sequence of local sampling instants $\bm{\tau}=\{\tau_k\}_{k=0}^\infty$, adopt the notation $\mf{x}[k]:=\mf{x}(\tau_k)$ and note that a sampled-data version of the target dynamics \eqref{eq:target} for $\mf{x}(t)$ is obtained similarly as in \cite[Section 4.5.2]{Soderstrom}:
\begin{equation}
\label{eq:discrete}
    \mf{x}(t) = \mf{A}_{\mfs{d}}(t-\tau_k)\mf{x}[k] + \mf{u}_{\mfs{d}}(t), \ \ t\in[\tau_k, \tau_{k+1}]
\end{equation}
where $\mf{u}_{\mfs{d}}(t)$ is a normal random variable of zero mean and  $\cov\{\mf{u}_{\mfs{d}}(t)\}=\mf{W}_{\mfs{d}}(t-\tau_k)=\int_{0}^{t} \mf{A}^{\mfs{d}}(\tau)\mf{BWB}^\top \mf{A}^{\mfs{d}}(\tau)^\top \text{\normalfont d} \tau$ with $\mf{A}^{\mfs{d}}(\tau) = \exp(\mf{A}\tau)$.

Every robot can use its observations of the target $\mf{z}[0],\dots,\mf{z}[k-1]$ and compute a causal optimal filter for \eqref{eq:discrete} at the sampling instants $t\in\bm{\tau}$ of the form $\hat{\mf{x}}^*[k]=\mathbb{E}\{\mf{x}[k]|\mf{z}[0],\dots,\mf{z}[k-1]\}$ which does not take into account $\mf{z}[k]$ due to the perception latency. By applying \cite[Page 228 - Theorem 4.1]{astrom} to system \eqref{eq:discrete} evaluated at $t=\tau_{k+1}$, the recursive filter structure to obtain $\hat{\mf{x}}^*[k]$ and $\hat{\mf{P}}^*[k]=\cov\{{\mf{x}}[k]-\hat{\mf{x}}^*[k]\}$ has the following form:
\begin{equation}
\label{eq:kalman}
    \begin{aligned}
    \mf{L}[k]&=\mf{A}_{\mfs{d}}(\Delta_k)\hat{\mf{P}}^*[k]\mf{C}^\top\left(\mf{C}\hat{\mf{P}}^*[k]\mf{C}^\top+\mf{R}(\Delta_k)\right)^{-1} \\
    \hat{\mf{x}}^*[k+1] &= \mf{A}_\mfs{d}(\Delta_k)\hat{\mf{x}}^*[k]+\mf{L}[k]\left(\mf{z}[k]-\mf{C}\hat{\mf{x}}^*[k]\right)\\
    \hat{\mf{P}}^*[k+1] &=\left(\mf{A}_\mfs{d}(\Delta_k)-\mf{L}[k]\mf{C}\right)\hat{\mf{P}}^*[k]\left(\mf{A}_\mfs{d}(\Delta_k)-\mf{L}[k]\mf{C}\right)^\top\\&+\mf{L}[k]\mf{R}(\Delta_k)\mf{L}[k]^\top+\mf{W}_\mfs{d}(\Delta_k)\\
    \end{aligned}
\end{equation}
assuming knowledge of some initial values for $\hat{\mf{x}}^*[0], \hat{\mf{P}}^*[0]$. On the other hand, an optimal estimation $\hat{\mf{x}}^*(t|k):=\mathbb{E}\{\mf{x}(t)|\mf{z}[0],\dots,\mf{z}[k-1]\}$ for $t\in(\tau_k,\tau_{k+1})$ can be obtained by using a model based prediction of \eqref{eq:discrete} as:
\begin{equation}
\label{eq:estimate_time}
\begin{aligned}
\hat{\mf{x}}^*(t|k) &=  \mf{A}_\mfs{d}(t-\tau_k)\hat{\mf{x}}^*[k]\\
\hat{\mf{P}}^*(t|k) &= \cov\{\mf{x}(t)-\hat{\mf{x}}^*(t|k)\} \\&= \mf{A}_\mfs{d}(t-\tau_k)\hat{\mf{P}}^*[k]\mf{A}_\mfs{d}(t-\tau_k)^T + \mf{W}_\mfs{d}(t-\tau_k)
\end{aligned}
\end{equation}
Note that the expressions in \eqref{eq:estimate_time} can be defined for all $t\geq 0$. However, we use $\hat{\mf{x}}^*(t)$ to refer to the whole optimal causal estimated trajectory constructed piecewise as $\hat{\mf{x}}^*(t):=\hat{\mf{x}}^*(t|k)$ and $\hat{\mf{P}}^*(t):=\hat{\mf{P}}^*(t|k)$ for $t\in[\tau_k,\tau_{k+1})$. \RevOne{Note also that \eqref{eq:kalman} is a Kalman filter \cite[Chapter 7]{astrom}. Hence, the standard discrete-time Kalman filter coincides with the DOE at $t=\tau_k$.}
\begin{rem}
\label{rem:doe}
Note that $\hat{\mf{x}}^*(t)$ is a smooth function of time for all $t\notin \bm{\tau}$. However, $\hat{\mf{x}}^*(t)$ may be discontinuous at $t\in\bm{\tau}$ due to the measurement corrections performed in \eqref{eq:kalman}. Hence we refer to $\hat{\mf{x}}^*(t)$ as a Discontinuous Optimal Estimation (DOE). If $\hat{\mf{x}}^*(t)$ is used to construct a reference for the robot, such discontinuities can cause persistent transients in the closed loop behaviour of the robot compromising asymptotic stability. An example of this is shown in Section \ref{sec:single}. 
\end{rem}
We propose a near-optimal alternative estimation $\hat{\mf{x}}(t)$ which does not have the discontinuity issue at $t\in\bm{\tau}$. The idea is to combine the current optimal estimate $\hat{\mf{x}}^*(t|k)$ for $t\in[\tau_k, \tau_{k+1})$ with the continued prediction from the previous estimate $\hat{\mf{x}}^*(t|k-1)$ for $t\geq \tau_k$, using a smooth transition between both. 

In order to perform such transition, consider the following auxiliary function:

\begin{defn}[Transition function] \label{def:mod}  $\upeta(\bullet;\alpha):[0,1]\to\mathbb{R}$ is a transition function if it complies that $\forall\mu\in\{1,\dots,m\}$:
\begin{enumerate}[label={\normalfont\textbf{\roman*})}]
    \item\label{eta:continous} $\upeta^{(\mu)}\left(t;\alpha\right)$ exists $\forall t\in[0,1]$.
    \item\label{eta:derivatives} $\upeta^{(\mu)}\left(0;\alpha\right) = \upeta^{(\mu)}\left(1;\alpha\right) = 0$.
    \item\label{eta:ends} $\upeta\left(0;\alpha\right) = 0$, $\upeta\left(1;\alpha\right)=1$.
    \item\label{eta:pointwise} $\lim_{\alpha\to 0}\upeta(t,\alpha)= 1, \forall t\in(0,1]$
\end{enumerate}
\end{defn}
An example of a transition function is provided in \ref{ap:mod} which can be assumed to be fixed as such through the rest of the manuscript. Hence, given $\alpha>0$, a transition function, and by defining the information matrix $\hat{\mf{Q}}^{*}(t|k)=\hat{\mf{P}}^{*}(t|k)^{-1}$ and information vector $\hat{\mf{y}}^{*}(t|k)=\hat{\mf{Q}}^{*}(t|k)\hat{\mf{x}}^{*}(t|k)$, let the Smooth Output Estimation (SOE):
\begin{equation}
\begin{aligned}
\label{eq:estimate}
    \uplambda_1(t|k)&= 1-\upeta((t-\tau_k)/\Delta_k;\alpha)\\
    \uplambda_2(t|k)&= \upeta((t-\tau_k)/\Delta_k;\alpha)\\
    \hat{\mf{Q}}(t) &= \uplambda_1(t|k)\hat{\mf{Q}}^*(t|k-1) + \uplambda_2(t|k)\hat{\mf{Q}}^*(t|k)\\
    \hat{\mf{y}}(t)&=  \uplambda_1(t|k)\hat{\mf{y}}^*(t|k-1) + \uplambda_2(t|k)\hat{\mf{y}}^*(t|k) \\
    \hat{\mf{P}}(t)&=\hat{\mf{Q}}(t)^{-1}\\
    \hat{\mf{x}}(t)&=\hat{\mf{Q}}(t)^{-1}\hat{\mf{y}}(t)
\end{aligned}
\end{equation}
for $t\in[\tau_k, \tau_{k+1})$. Similarly as with $\hat{\mf{x}}^*(t)$, we drop the dependence on $k$ in the notation for the SOE estimation $\hat{\mf{x}}(t)$ in order to refer to the whole sub-optimal trajectory. Now, we establish some important properties of \eqref{eq:estimate}.
\begin{thm}
\label{th:smooth}
Given $\alpha>0$, the SOE in \eqref{eq:estimate} complies with the following:
\begin{enumerate}[label={\normalfont\textbf{\roman*})}]
\item {\normalfont\textbf{(Smoothness) }} The information vector and matrix $\hat{\mf{y}}(t), \hat{\mf{Q}}(t)$ are $m$-times differentiable $\forall t\geq 0$. As a consequence, $\hat{\mf{x}}(t)$ is $m$-times differentiable $\forall t\geq 0$.
\item {\normalfont\textbf{(Unbiasedness) }} $\mathbb{E}\{\mf{x}(t)-\hat{\mf{x}}(t)\}=0, \forall t\geq 0$.
\item {\normalfont\textbf{(Tight consistency) }} 
    $\hat{\mf{P}}^*(t) \preceq\cov\{\mf{x}(t)-\hat{\mf{x}}(t)\} \preceq\hat{\mf{P}}(t), \forall t\geq 0$ with equality for all $t\notin\bm{\tau}$ as $\alpha\to 0$.
\end{enumerate}
\end{thm}
\begin{pf}
The proof can be found in \ref{ap:smooth}.
\end{pf}
\begin{rem}
\label{rem:smooth}
\RevOne{Different from a discrete-time Kalman filter or the DOE, the SOE is ensured to have a sufficiently smooth output for any $t\geq 0$, even under large perception latency.} There are two basic smooth-output alternatives to the SOE proposed in \eqref{eq:estimate}: interpolation techniques and low-pass filtering. In the case of interpolation, an $m$-th order spline may be used in a moving horizon fashion to fill in the spaces between samples $\{\hat{\mf{y}}^*[k],\hat{\mf{Q}}^*[k]\}_{k=0}^\infty$ for all $t\notin\bm{\tau}$. On the other hand, an $m$-th order low-pass filter can be used on top of \eqref{eq:estimate_time} to obtain smooth versions of $\hat{\mf{y}}^*(t),\hat{\mf{Q}}^*(t)$ as well. Therefore, an analogous to property \textbf{i}) of Theorem \ref{th:smooth} is obtained for both of these alternatives. However, neither unbiasedness nor consistency can be ensured for such types of estimations. In addition, contrary to the alternatives, the tight near-optimal quality of the estimation of our proposal can be adjusted appropriately by modifying the parameter $\alpha>0$ as shown in item \textbf{iii}) of Theorem \ref{th:smooth}.
\end{rem}

\begin{rem}
\label{rem:false}
\RevTwo{In practice, VTDs such as the ones in \cite{redmon2016,ren2015,Dai2016} are prone to data association problems when false positives or miss-detections occur, and under false negatives and occlusions where the procedure cannot detect the target. However, a VTD in conjunction with target tracking using state-based predictions provide a more robust alternative since predicted targets can be matched with detections. For instance, \cite{song2019} provides some examples in which state-based predictions are used to improve the accuracy of the target detector. In this context, the SOE can be used as a state-based prediction by evaluating \eqref{eq:estimate} at any time until a new sample is available, which can then be used for data association.}
\end{rem}

\section{Estimation fusion}
\label{sec:fusion}
Using the SOE in \eqref{eq:estimate}, each robot $i$ is able to compute a near-optimal local estimation $\hat{\mf{x}}_i(t)$ and a consistent covariance matrix $\hat{\mf{P}}_i(t)$ with their corresponding smooth information vector and matrix $\hat{\mf{y}}_i(t), \mf{\hat{Q}}_i(t)$ as a result of Theorem \ref{th:smooth}. Hence, we turn our attention to the task of fusing this information across all  robots in the communication network. Note that even when processed measurements are uncorrelated, the estimations $\hat{\mf{x}}_i(t)$ are correlated since both depend on the same noise process $\mf{u}(t)$ driving system \eqref{eq:target}. However, keeping track of the cross-correlations between estimates at different robots is not scalable with respect to the network size. Therefore, we ignore cross-correlations and fuse estimation information according to:
\begin{equation}
\begin{aligned}
\mf{\bar{Q}}_\mfs{G}(t) &= \dfrac{1}{\mfs{N}}\sum_{i=1}^\mfs{N}\mf{\hat{Q}}_i(t), &\ \  \mf{\bar{y}}_\mfs{G}(t)&=\dfrac{1}{\mfs{N}}\sum_{i=1}^\mfs{N}\mf{\hat{y}}_i(t) \\[0.5em]
\bar{\mf{x}}_\mfs{G}(t) &= \mf{\bar{Q}}_\mfs{G}(t)^{-1}\bar{\mf{y}}_\mfs{G}(t), & \bar{\mf{p}}_\mfs{G}(t)&=\mf{C}\bar{\mf{x}}_\mfs{G}(t) \label{eq:output}
\end{aligned}
\end{equation}
Note that the matrix $\bar{\mf{P}}_{\mfs{G}}(t) = \bar{\mf{Q}}_{\mfs{G}}(t)^{-1}$ is consistent in the sense that $\cov\{{\mf{x}}(t) - \bar{\mf{x}}_\mfs{G}(t)\}\preceq \bar{\mf{P}}_{\mfs{G}}(t)$ since $\bar{\mf{Q}}_{\mfs{G}}(t)$ is a convex combination of the information matrices $\{\hat{\mf{Q}}_i(t)\}_{i=1}^\mfs{N}$ and by the covariance-intersection principle from \cite[Section 2.1]{cov_intersection}. The same idea of computing a consistent inverse-covariance fusion estimate as in \eqref{eq:output}, ignoring cross-correlations, has shown to be very successfully in different state estimation contexts such as \cite{kalman_ros}.

If every robot had access to the global quantity $\bar{\mf{p}}_{\mfs{G}}(t)$ from \eqref{eq:output}, hence a local trajectory ${\mf{p}}^{\mfs{r}}_i(t)=\bar{\mf{p}}_{\mfs{G}}(t)+\mf{d}_i$ can be constructed for each agent to follow. However, in order to construct a controller $\mf{u}_i(t)$ to achieve trajectory tracking of \eqref{eq:agents} towards ${\mf{p}}^{\mfs{r}}_i(t)$, knowledge of the derivatives $({\mf{p}}^{\mfs{r}}_i)^{(\mu)}(t)=\bar{\mf{p}}_{\mfs{G}}^{(\mu)}(t), \mu\in\{1,\dots,m\}$ is required as we discuss in Section \ref{sec:control}. While the expressions in \eqref{eq:output} constitute a transformation between information and state representations, the following result provides an equivalent transformation between the derivatives of $\bar{\mf{p}}_{\mfs{G}}(t)$ and the ones of $\bar{\mf{Q}}_{\mfs{G}}(t),\bar{\mf{y}}_{\mfs{G}}(t)$.

\begin{lem}
\label{le:derivatives}
Let $\bar{\mf{p}}_{\mfs{G}}(t). \bar{\mf{Q}}_{\mfs{G}}(t), \bar{\mf{y}}_{\mfs{G}}(t)$ defined in \eqref{eq:output} and $\bar{\mf{P}}_{\mfs{G}}(t)=\bar{\mf{Q}}_{\mfs{G}}(t)^{-1}$. Then, for any $\mu\in\{1,\dots,m\}$:
\begin{enumerate}[label={\normalfont\textbf{\roman*})}]
\item $\bar{\mf{P}}_{\mfs{G}}^{(\mu)}(t) =- \bar{\mf{P}}_{\mfs{G}}(t)\sum_{\nu=0}^{\mu-1}\binom{\mu}{\nu}\bar{\mf{Q}}_{\mfs{G}}^{(\mu-\nu)}(t)\bar{\mf{P}}_{\mfs{G}}^{(\nu)}(t)$.
\item $\bar{\mf{p}}_{\mfs{G}}^{(\mu)}(t) =\mf{C} \sum_{\nu=0}^\mu\binom{\mu}{\nu}\bar{\mf{P}}_{\mfs{G}}^{(\nu)}(t)\bar{\mf{y}}_{\mfs{G}}^{(\mu-\nu)}(t)$.
\end{enumerate}
\end{lem}
\begin{pf}
The proof can be found in \ref{ap:derivatives}.
\end{pf}

Computing \eqref{eq:output}  as well as the \RevAll{derivatives} in Lemma \ref{le:derivatives}-\textbf{ii}) in a decentralized fashion under time-varying $\mf{\hat{y}}_i(t), \mf{\hat{Q}}_i(t)$ is not trivial. In the following we provide an algorithm to compute such quantities asymptotically without steady state error using exact dynamic consensus tools, even under persistently varying $\mf{\hat{y}}_i(t), \mf{\hat{Q}}_i(t)$. The fusion algorithm is composed by two sequential stages. First, we use a \textit{consensus stage} where information fusion is performed to compute $\bar{\mf{y}}_{\mfs{G}}(t),\bar{\mf{Q}}_{\mfs{G}}(t)$ and its first $m$ derivatives using local information  $\{\hat{\mf{y}}_i(t),\hat{\mf{Q}}_i(t)\}_{i=1}^\mfs{N}$ in a decentralized fashion. Second, we use an \textit{output stage} where the outputs of the consensus stage are organized in order to compute $\bar{\mf{p}}_{\mfs{G}}(t)$ and its first $m$ derivatives based on the transformations given in \eqref{eq:output} and Lemma \ref{le:derivatives}.

Based on the REDCHO algorithm \cite{redcho} given in \ref{ap:redcho_prot} with parameters $\{k_\mu,\gamma_\mu\}_{\mu=0}^m$ and $\theta$, the consensus stage is composed by  consensus protocols for the information vector
\begin{equation}
\label{eq:redcho_vector}
\begin{array}{ll}
    \mf{y}_{i,\mu}(t) &= \hat{\mf{y}}_i^{(\mu)}(t) - \sum_{\nu=0}^{m} G_{\mu+1,\nu+1} {\mf{v}}_{i,\nu}(t)\\
    \dot{{\mf{v}}}_{i,\mu}(t) &= k_\mu\theta^{\mu+1} \sum_{j=1}^\mfs{N}a_{ij}\lceil \mf{y}_{i,0}(t) - \mf{y}_{j,0}(t) \rfloor^{\frac{m-\mu}{m+1}}+\ \mf{v}_{i,\mu+1}(t) - \gamma_\mu \mf{v}_{i,\mu}(t), \\ & \text{for } 0\leq \mu < m\\
    \dot{{\mf{v}}}_{i,m}(t) &= k_m\theta^{m+1} \sum_{j=1}^\mfs{N}a_{ij}\sgn{\mf{y}_{i,0}(t) - {\mf{y}}_{j,0}(t) }{0} - \gamma_m{\mf{v}}_{i,m}(t)
\end{array}
\end{equation}
and for the information matrix
\begin{equation}
\label{eq:redcho_matrix}
\begin{array}{ll}
    \mf{Q}_{i,\mu}(t) &= \hat{\mf{Q}}_i^{(\mu)}(t) - \sum_{\nu=0}^{m} G_{\mu+1,\nu+1} {\mf{V}}_{i,\nu}(t)\\
    \dot{{\mf{V}}}_{i,\mu}(t) &= k_\mu\theta^{\mu+1} \sum_{j=1}^\mfs{N}a_{ij}\lceil \mf{Q}_{i,0}(t) - \mf{Q}_{j,0}(t) \rfloor^{\frac{m-\mu}{m+1}}+\ \mf{V}_{i,\mu+1}(t) - \gamma_\mu \mf{V}_{i,\mu}(t), \\ & \text{for } 0\leq \mu < m\\
    \dot{{\mf{V}}}_{i,m}(t) &= k_m\theta^{m+1} \sum_{j=1}^\mfs{N}a_{ij}\sgn{\mf{Q}_{i,0}(t) - {\mf{Q}}_{j,0}(t) }{0} - \gamma_m{\mf{V}}_{i,m}(t)
\end{array}
\end{equation}
where $a_{ij}$ are the components of the adjacency matrix of $\mathcal{G}$, $G_{\mu+1,\nu+1}$ with $\mu,\nu\in\{0,\dots,m\}$ are the components of $\mf{G}$ defined in \ref{ap:aux_mat}.

The protocols in \eqref{eq:redcho_vector} and \eqref{eq:redcho_matrix} take as an input the local information $\hat{\mf{y}}_i(t),\hat{\mf{Q}}_i(t)$, and have outputs $\{\mf{y}_{i,\mu}(t),\mf{Q}_{i,\mu}(t)\}_{\mu=0}^m$ computed \RevAll{through} the internal variables $\{\mf{v}_{i,\mu}(t),\mf{V}_{i,\mu}(t)\}_{\mu=0}^m$. The purpose of \eqref{eq:redcho_vector} is that $\mf{y}_{i,0}(t),$ $\dots,\mf{y}_{i,m}(t)$ will converge towards $\bar{\mf{y}}_{\mfs{G}}(t)$ and its first $m$ derivatives for each robot $i\in\{1,\dots,\mfs{N}\}$. Similarly, $\mf{Q}_{i,0}(t),\dots, \mf{Q}_{i,m}(t)$ will converge towards $\bar{\mf{Q}}_{\mfs{G}}(t)$ and its first $m$ derivatives. In addition, the structure of the protocol allows each agent to communicate only $\mf{y}_{i,0}(t)$ and $\mf{Q}_{i,0}(t)$ to its neighbors. It is clear that each robot requires to compute $\{\hat{\mf{y}}^{(\mu)}_i(t), \hat{\mf{Q}}_{i}^{(\mu)}(t)\}_{\mu=0}^m$ locally as observed in equations for $\mf{y}_{i,\mu}(t), \mf{Q}_{i,\mu}(t)$ in \eqref{eq:redcho_vector} and \eqref{eq:redcho_matrix}. These can be computed explicitly from \eqref{eq:estimate} since the expression for the transition function $\upeta(\bullet;\alpha)$ and the model based predictions in \eqref{eq:estimate_time} are known explicitly as well.

Moreover, note that each component of the equations in \eqref{eq:redcho_vector} and \eqref{eq:redcho_matrix} is an independent instance of the REDCHO protocol in \eqref{eq:redcho} from \ref{ap:redcho_prot}. As a result, \eqref{eq:redcho_vector} and \eqref{eq:redcho_matrix} can be alternatively implemented using $nm + nm(nm+1)/2$ REDCHO instances, one for each non-repeated component of $\mf{y}_{i,m}(t)\in\mathbb{R}^{nm},\mf{Q}_{i,m}(t)\in\mathbb{R}^{nm\times nm}$ provided that $\mf{Q}_{i,m}(t)$ is symmetric.

In the output stage, $m+1$ outputs are obtained from $\{\mf{y}_{i,\mu}(t),\mf{Q}_{i,\mu}(t)\}_{\mu=0}^m$ as given in Algorithm \ref{algo:spatial} which is based on the transformation of Lemma \ref{le:derivatives}. In fact, the structure in Algorithm \ref{algo:spatial} is equivalent to the one in Lemma~\ref{le:derivatives} assuming that $\{\mf{y}_{i,\mu}(t),\mf{Q}_{i,\mu}(t)\}_{\mu=0}^m$ have already converged towards $\bar{\mf{y}}_{\mfs{G}}(t),\bar{\mf{Q}}_{\mfs{G}}(t)$ and their first $m$ derivatives. Hence, in order to ensure \RevAll{convergence} we adopt the following assumption.
\begin{algorithm}
  \begin{algorithmic}[1]
\State \texttt{inputs:}  $\{\mf{y}_{i,\mu}(t),\mf{Q}_{i,\mu}(t)\}_{\mu=0}^m$ \texttt{ computed from \eqref{eq:redcho_vector} and \eqref{eq:redcho_matrix}}
\State $\mf{P}_{i,0}(t) \leftarrow \mf{Q}_{i,0}(t)^{-1}$
  \State $\mf{p}_{i,0}(t)\leftarrow \mf{C}\mf{P}_{i,0}(t)\mf{y}_{i,0}(t)$
  \For {$\mu\in\{1,\dots,m\}$} 
  \State $\mf{P}_{i,\mu}(t)\leftarrow -\mf{P}_{i,0}(t)\sum_{\nu=0}^{\mu-1}\binom{\mu}{\nu}\mf{Q}_{i, \mu - \nu}(t)\mf{P}_{i,\nu}(t)$ 
  \State $\mf{p}_{i,\mu}(t)\leftarrow \mf{C}\sum_{\nu=0}^{\mu}\binom{\mu}{\nu}\mf{P}_{i,\nu}(t)\mf{y}_{i,\mu-\nu}(t)$
  \EndFor
  \State \texttt{return }$\{\mf{p}_{i,\mu}(t)\}_{\mu=0}^m$

 \caption{Estimation fusion output stage}\label{algo:spatial}
\end{algorithmic}
\end{algorithm}
\begin{assum}
\label{as:consensus}
Let gains $\gamma_0,\dots,\gamma_m>0$ and $\hat{s}_i(t)$ be an arbitrary component of $\mf{\hat{y}}_i(t)$ or $\mf{\hat{Q}}_i(t)$. Hence, $\{\hat{s}_i(t)\}_{i=1}^{\mfs{N}}$ satisfy Assumption \ref{as:u_signals} in \ref{ap:redcho_prot} for some $L>0$.
\end{assum}
\begin{rem}
Due to item \textbf{i)} of Theorem \ref{th:smooth},  Assumption \ref{as:consensus} is complied for $t$ on any compact interval and sufficiently big $L>0$. However, this assumption can be complied for any $t\geq 0$ as well by assuming that the motion of the target has bounded derivatives, which is mild in practical scenarios.
\end{rem}

\begin{thm}
\label{th:spatial}
Let Assumption \ref{as:consensus} hold and $\mathcal{G}$ be a connected network. Moreover, let the initial conditions and gains for each REDCHO instance configured to satisfy the conditions of Proposition \ref{prop:redcho} in \ref{ap:redcho_prot}. Then, there exists an $m$-times differentiable consensus signal $\mf{\tilde{p}}(t)\in\mathbb{R}^n$ and $T>0$ such that the outputs of Algorithm \ref{algo:spatial} satisfy that for all $\mu\in\{0,\dots,m\}$:
$$
\mf{\tilde{p}}^{(\mu)}(t)=\mf{p}_{1,\mu}(t)=\cdots=\mf{p}_{\mfs{N},\mu}(t), \forall t\geq T
$$
In addition, $\lim_{t\to\infty}\left|\mf{\tilde{p}}^{(\mu)}(t)- \mf{\bar{p}}_\mfs{G}^{(\mu)}(t)\right| = 0$ 

\end{thm}
\begin{pf}
The proof can be found in \ref{ap:spatial}.
\end{pf}
\begin{rem}
The interpretation of the convergence result in Theorem \ref{th:spatial} is that the outputs of Algorithm \ref{algo:spatial} for all robots will converge to some arbitrary consensus signal $\tilde{\mf{p}}(t)$ in finite time, which is not necessarily the average $\bar{\mf{p}}_{\mfs{G}}(t)$. Achieving consensus in finite time is interesting from the point of view of formation control since it allows to the observer and the controller to be designed independently without compromising the stability of the overall system as discussed in Section \ref{sec:control}. Moreover, all robots will place themselves around the same formation center $\tilde{\mf{p}}(t)$ in finite time, which will converge towards $\bar{\mf{p}}_{\mfs{G}}(t)$ according to Theorem \ref{th:spatial}.
\end{rem}

\section{Formation control}
\label{sec:control}
Equipping all agents with the previously presented tools for distributed estimation we construct a local controller of the form:

\begin{equation}
\label{eq:control}
\begin{aligned}
    \mf{u}_i(t) &= \mf{p}_{i,m}(t) - \kappa_0(\mf{p}_i(t) - \mf{p}_{i,0}(t) - \mf{d}_i)   -\sum_{\mu=1}^{m-1} \kappa_\mu(\mf{p}^{(\mu)}_i(t)- \mf{p}_{i,\mu}(t))
\end{aligned}
\end{equation}

where the roots of the polynomial $\lambda^m + \sum_{\mu=0}^{m-1}\kappa_\mu \lambda^{\mu}$ have negative real part.
\begin{prop}
\label{prop:control}
Let the conditions of Theorem \ref{th:spatial} hold. Define, $\mf{p}_i^{\mfs{r}}(t):=\bar{\mf{p}}_{\mfs{G}}(t) + \mf{d}_i$ with $\bar{\mf{p}}_{\mfs{G}}(t)$ in \eqref{eq:output}. Then, the closed loop system \eqref{eq:agents} under controller \eqref{eq:control} satisfies the formation goal in \eqref{eq:goal} regardless of $\mf{x}_i(0)$.
\end{prop}
\begin{pf}
First, note that due to Theorem \ref{th:spatial}, then there exists $T>0$ such that $\mf{p}_{i,\mu}(t)\equiv \tilde{\mf{p}}^{(\mu)}(t)$. Moreover, the outputs $\mf{p}_{i,\mu}(t)$ remain bounded for $t\in[0,T]$. The closed loop system \eqref{eq:agents} under \eqref{eq:control} is input to state stable with respect to $\mf{u}_i(t)$. Thus, there are no finite-time escapes of $\mf{x}_i(t)$ for any $t\in[0,T]$. For $t\geq T$, \eqref{eq:control} takes the form

\begin{equation}
\begin{aligned}
    \mf{u}_i(t) &= \tilde{\mf{p}}^{(m)}(t) - \kappa_0(\mf{p}_i(t) - \tilde{\mf{p}}(t) - \mf{d}_i)  
     -\sum_{\mu=1}^{m-1} \kappa_\mu(\mf{p}_i^{(\mu)}(t)- \tilde{\mf{p}}^{(\mu)}(t))
\end{aligned}
\end{equation}

Now, define $\mf{e}(t)=\mf{p}_i(t) - \tilde{\mf{p}}(t) - \mf{d}_i$ to obtain $\mf{e}^{(m)}(t) = - \sum_{\mu=0}^{m-1}\kappa_\mu \mf{e}^{(\mu)}(t)$ which is asymptotically stable towards the origin. Hence, $\mf{p}_i(t)$ converge asymptotically towards $\tilde{\mf{p}}(t)+\mf{d}_i$ which converge towards $\bar{\mf{p}}_\mfs{G}(t)+\mf{d}_i$ due to Theorem \ref{th:spatial}, implying \eqref{eq:goal}.
\end{pf}
\begin{rem}
Note that a linear controller was proposed in \eqref{eq:control}, whose design is decoupled from the sampling instants $\bm{\tau}_i$ arising from the perception mechanism. As evident from the proof of Proposition \ref{prop:control} and since the outputs of the estimation fusion technique from Algorithm \ref{algo:spatial} converge all $m+1$ derivatives of $\mf{\bar{p}}_{\mfs{G}}(t)$, then these ideas can be extended directly to other trajectory tracking controllers, without requiring a co-design between the perception mechanism and the controller in order to ensure stability.
\end{rem}

\section{Simulation examples}
\label{sec:simulations}
For simplicity in the presentation, we assume that each of the $n$ coordinates is uncorrelated both in the detection and in the target model \eqref{eq:target}. Hence, it suffices to analyse each coordinate by separate. Equivalently, we consider $n=1$. Now, let $m=2$ for \eqref{eq:agents} and \eqref{eq:target} to represent second-order integrators. All REDCHO instances are configured with $m=2$ and gains $k_0=6, k_1=11, k_2=6, \gamma_0=\gamma_1=\gamma_2=1, \theta=40$ obtained using the tuning rules from \cite{edcho,redcho}. Similarly, all robots use the controller \eqref{eq:control} with $\kappa_0=1, \kappa_1=2$. 

\RevTwo{As characterized by \cite{huang2017}, it is expected that a standard target detector based on, e.g., convolutional neural networks, improve its performance when more computing power is employed, effectively increasing its perception latency. Hence, to illustrate the performance of our proposal under this perception latency and detection quality trade-off, the perception mechanism is configured with two possible perception methods with latencies $\Delta^1=1,\Delta^2=0.5$. These correspond to a computation-intensive detection method and a lighter one, respectively. As a result, for illustration purposes, we choose
covariance matrices $\mf{R}^1=0.01, \mf{R}^2=0.1$ respectively to simulate that the first method performs better than the second one at the expense of large computing time.}

Furtheremore, let $\mf{W}=1$ for the noise process in \eqref{eq:target}. The stochastic system \eqref{eq:target} was simulated using Euler-Maruyama with time step $\Delta t=10^{-6}$ whereas the REDCHO instances in \eqref{eq:redcho} from \ref{ap:redcho_prot} were simulated using explicit Euler method with the same time step.
\subsection{Single robot}
\label{sec:single}
\begin{figure}
\centering
    \includegraphics[width=0.5\textwidth]{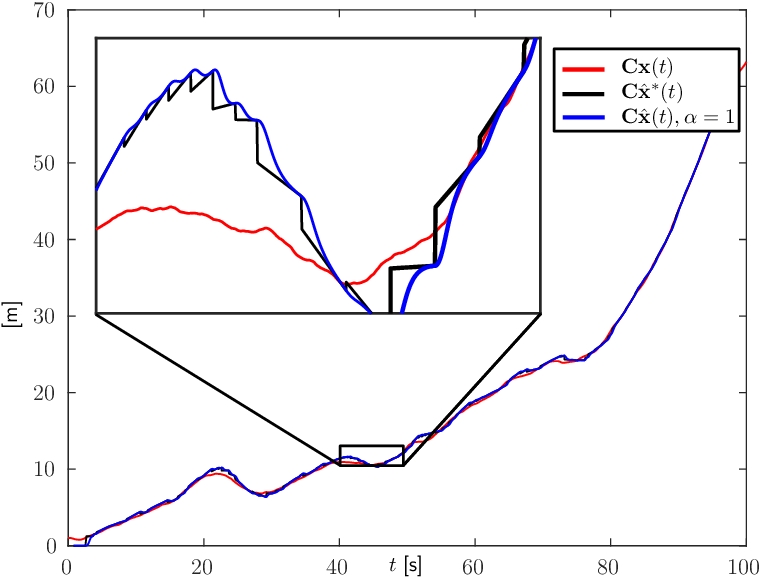}

     \caption{A realization of the target position $\mf{C}\mf{x}(t)$ as well as its corresponding estimations $\mf{C}\hat{\mf{x}}(t)$ and $\mf{C}\hat{\mf{x}}^*(t)$ for the SOE and the DOE respectively. Note that the SOE is always a smooth estimation whereas the DOE is discontinuous at the sampling instants.}
    \label{fig:kalman_output}
\end{figure}

\begin{figure}
\centering

    \includegraphics[width=0.5\textwidth]{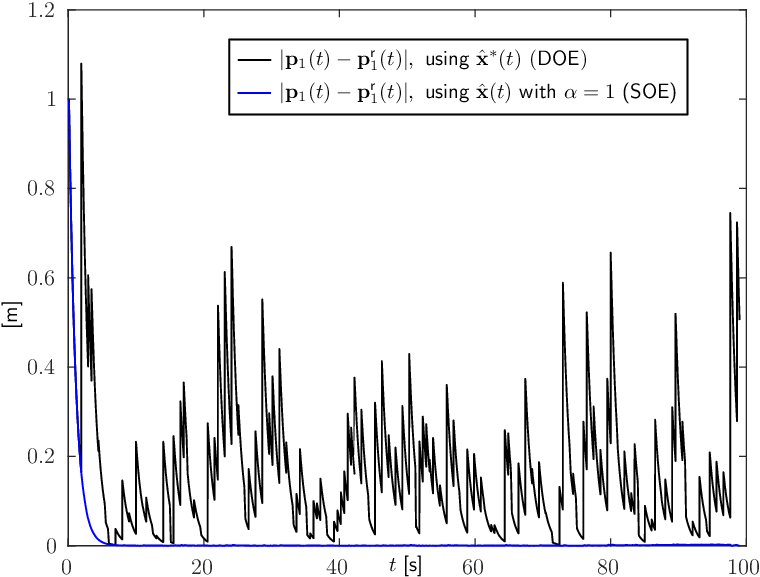}

     \caption{Trajectory tracking performance of a single robot \eqref{eq:agents} under the control input \eqref{eq:local_control}. Note that the reference is always smooth when using the SOE, resulting in asymptotic convergence for the tracking error. However, the reference is discontinuous when using the DOE, which leads to persistent transients in the robot performance, preventing asymptotic convergence of the tracking error.}
    \label{fig:network}
\end{figure}

In this section, we consider $\mfs{N}=1$ in order to evaluate the performance of the SOE from Section \ref{sec:smooth_output}. Consider a randomly generated sequence of perception latencies $\{\Delta_{1,k}\}_{k=0}^{\infty}$ with $\Delta_{1,k}\in\{1,0.5\}$ leading to a sequence of sampling instants $\bm{\tau}_1=\{\tau_{1,k}\}_{k=1}^\infty$. Figure \ref{fig:kalman_output} shows a realization $\mf{x}(t)$ of \eqref{eq:target}, and the SOE $\hat{\mf{x}}_1(t)$ from \eqref{eq:estimate} for $\alpha=1$. \RevOne{Similarly, we show the output $\hat{\mf{x}}^*_1(t)$ for the DOE in \eqref{eq:estimate_time}  for comparison which coincides with the Kalman filter at $t\in\bm{\tau}_1$.} We construct a control input analogous to \eqref{eq:control} as:
\begin{equation}
\label{eq:local_control}
    \mf{u}_i(t) = \ddot{\mf{p}}_1^{\mfs{r}}(t) - \kappa_0(\mf{p}_1(t) - \mf{p}_1^{\mfs{r}}(t))-\kappa_1(\dot{\mf{p}}_1(t)-\dot{\mf{p}}_1^{\mfs{r}}(t))
\end{equation}
for all $t\notin\bm{\tau}_1$, where $\mf{p}_1^{\mfs{r}}(t)=\mf{C}\hat{\mf{x}}_1(t)+\mf{d}_1$ for the SOE and $\mf{p}_1^{\mfs{r}}(t)=\mf{C}\hat{\mf{x}}^*_1(t)+\mf{d}_1$ for the DOE. The expressions for the derivatives $ \ddot{\mf{p}}_1^{\mfs{r}}(t),\dot{\mf{p}}_1^{\mfs{r}}(t)$ can be obtained explicitly from \eqref{eq:estimate_time} and \eqref{eq:estimate} and are omitted here for brevity. Moreover, note that $ \ddot{\mf{p}}_1^{\mfs{r}}(t),\dot{\mf{p}}_1^{\mfs{r}}(t), \forall t\in\bm{\tau}_1$ does not exist when using the DOE.  Figure \ref{fig:network} shows the trajectory tracking performance of the robot \eqref{eq:agents} using the controller \eqref{eq:local_control} for each case. It can be observed that the tracking error converges to zero when using the SOE. On the other hand, persistent transients are observed when using the DOE due to the discontinuities in the reference at $t\in\bm{\tau}_1$.

\subsection{Multi-robot}
\label{sec:multi}
\begin{figure}
\centering

    \includegraphics[width=0.5\textwidth]{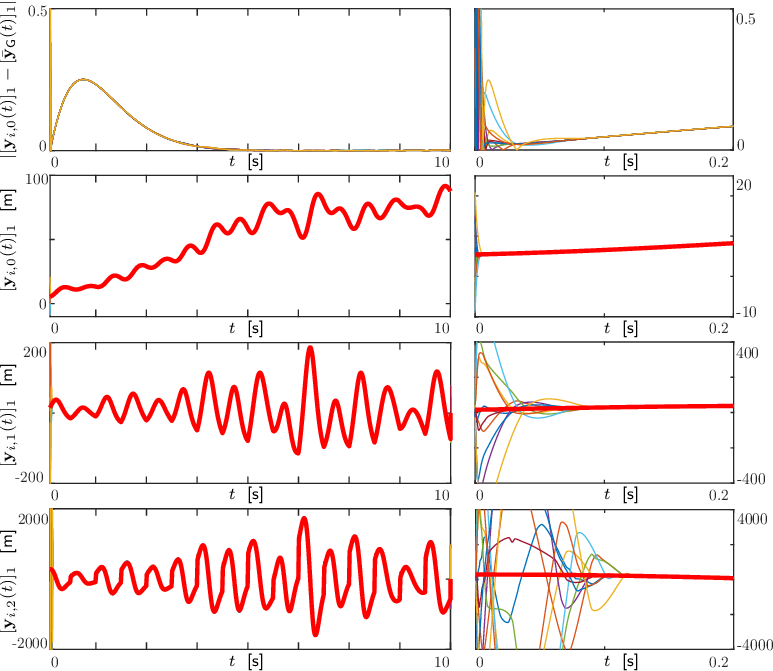}

     \caption{Trajectories for the first components of $\mf{y}_{i,0}(t), \mf{y}_{i,1}(t), \mf{y}_{i,2}(t)$ denoted as $[\mf{y}_{i,0}(t)]_1, [\mf{y}_{i,1}(t)]_1, [\mf{y}_{i,2}(t)]_1$, shown to converge to $[\bar{\mf{y}}_{\mfs{G}}(t)]_1,[\dot{\bar{\mf{y}}}_{\mfs{G}}(t)]_1,[\ddot{\bar{\mf{y}}}_{\mfs{G}}(t)]_1$ which appear in solid red color. Figures on the left show trajectories in the interval $t\in[0,10]$ to depict the asymptotic convergence behavior of the algorithm towards the global signals. On the other hand, figures on the right show convergence towards consensus which occurs in the interval \RevAll{$t\in[0,0.2]$}.  }
    \label{fig:information_convergence}
\end{figure}

\begin{figure}
\centering
    \includegraphics[width=0.45\textwidth]{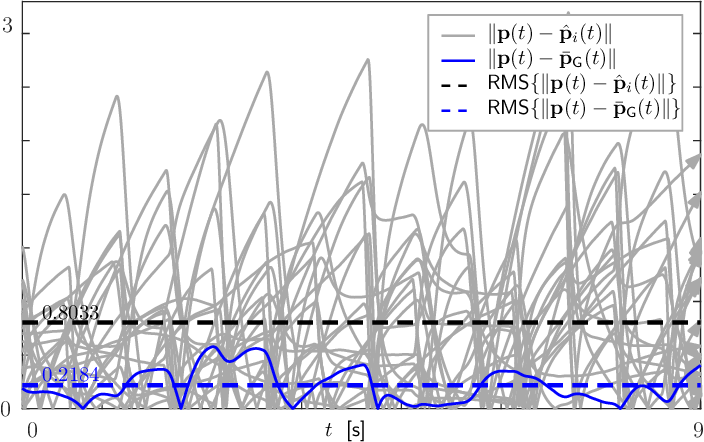}
     \caption{Error comparison between the actual realization of the target position  $\mf{p}(t)$, local SOE estimates $\hat{\mf{p}}_i(t)$ (grey) and the collaborative estimation $\bar{\mf{p}}_{\mfs{G}}(t)$ (blue). Moreover, average Root Mean Squared (RMS) values are shown in each case, where an improvement of a factor of $3.5$ is observed when comparing the collaborative estimate with respect to the local ones. }
    \label{fig:centralized_comparison}
\end{figure}

\begin{figure}
\centering

    \includegraphics[width=\textwidth]{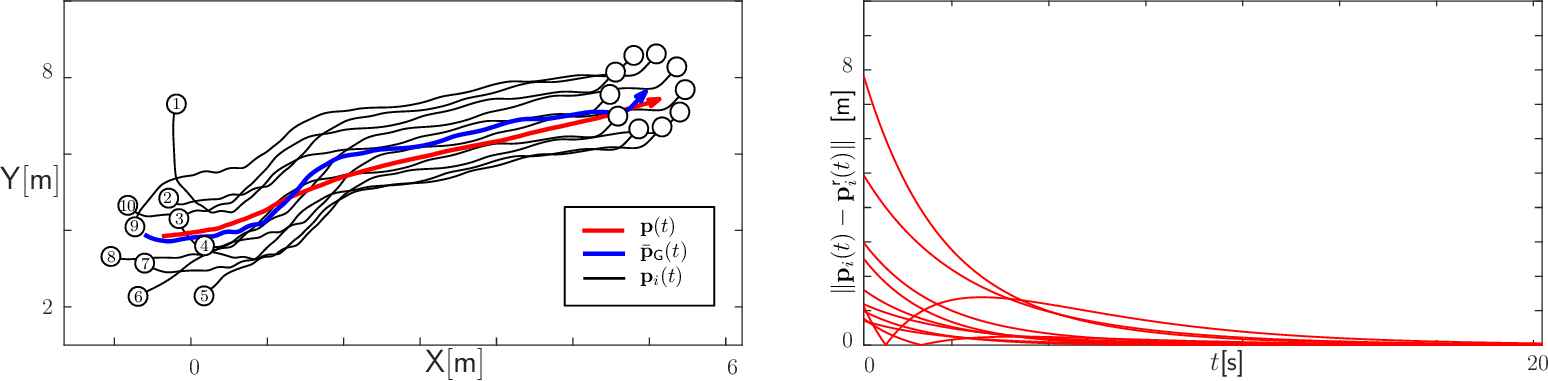}
     \caption{(Left) Actual target trajectory $\mf{p}(t)$ on the plane as well as the global estimation $\bar{\mathbf{p}}_{\mfs{G}}(t)$ and indiviual robot trajectories $\mathbf{p}_i(t)$ which converge to the circular formation around $\bar{\mathbf{p}}_{\mfs{G}}(t)$. (Right) Formation error between the robot position $\mathbf{p}_i(t)$ and the reference $\mathbf{p}_i^{\mfs{r}}(t)=\bar{\mf{p}}_{\mfs{G}}(t)+\mf{d}_i$.}

    \label{fig:formation}
\end{figure}

Now, consider a communication network of $\mfs{N}=10$ robots connected in a ring topology. The estimation fusion protocol was implemented by separate for the $\mfs{X}$ and $\mfs{Y}$ components of a two dimensional target trajectory $\mf{p}(t)$. Figure \ref{fig:information_convergence} shows the convergence of the first components of $\mf{y}_{i,0}(t),\mf{y}_{i,1}(t),\mf{y}_{i,2}(t)$ for the estimation of the $\mfs{X}$ coordinate, where it can be observed that all agents converge to a common signal in finite time, and then converge asymptotically to the first component of the centralized signal $\bar{\mf{y}}_{\mfs{G}}(t)$ and its derivatives. The convergence of the rest of the components of $\mf{y}_{i,0}(t),\mf{y}_{i,1}(t),\mf{y}_{i,2}(t)$ as well as for $\mf{Q}_{i,0}(t),\mf{Q}_{i,1}(t),\mf{Q}_{i,2}(t)$ is similar and is omitted here for the sake of brevity. \RevThree{In addition, we use the Root Mean Squared (RMS) error performance index in this experiment to measure the impact of using the fusion block. For a single experiment of duration $T$, the RMS value for an arbitrary scalar signal $x(t)$ is computed as
$$
\mfs{RMS}\{x(t)\} := \sqrt{\frac{1}{T}\int_0^Tx(t)^2\text{d}t}
$$
Furthermore, we compute the average RMS value for arbitrary scalar signals $\{x_i(t)\}_{i=1}^\mfs{N}$ as $\mfs{RMS}_\mfs{avg}\{x_i(t)\}_{i=0}^\mfs{N} := \frac{1}{\mfs{N}}\sum_{i=1}^\mfs{N}\mfs{RMS}\{x_i(t)\}$.}
The estimation error after the output stage of Algorithm \ref{algo:spatial} is shown in Figure \ref{fig:centralized_comparison} where it can be observed that the overall global estimate reduces the RMS error in a factor of 3.5 with respect to the average RMS of the individual local estimations. 
In addition, Figure \ref{fig:formation} shows the actual formation behaviour of the robots on the plane. Here, the robots start at random positions and converge to a circular formation around the target estimation $\bar{\mf{p}}_{\mfs{G}}(t)$.

\subsection{\RevOne{Ablation and parameter analysis}}
\label{sec:ablation}

\RevOne{In this section, we study the influence of the parameter $\alpha$ in the SOE as well as the fusion block.} \RevThree{For this purpose, we perform $\mfs{N}_{\mfs{MC}}=100$ Monte-Carlo runs, for each of the following configurations.} \RevOne{We use a similar setting as in Section \ref{sec:multi} with $\mfs{N}=10$ robots connected in a ring topology. We test the SOE with different values of the parameter $\alpha\in\{0.1,1,10\}$ as well as the DOE. In addition, for each value of $\alpha$, we test the performance of the system when using the fusion block or not for computation of the reference signal $\bar{\mf{p}}_\mfs{G}(t)$. In the case of the DOE, we do not include the fusion block, since the output of such estimator does not satisfy Assumption \ref{as:consensus} due to the discontinuities in the output, regardless of the motion of the target. When the fusion block is not used, we compute $\hat{\mf{p}}_i(t)=\mf{C}\hat{\mf{x}}_i(t)$ from the SOE \eqref{eq:estimate} or the DOE \eqref{eq:estimate_time} depending on the configuration. Otherwise, we compute $\hat{\mf{p}}_i(t)=\hat{\mf{p}}_{i,0}(t)$ as the output of the estimation fusion output stage from Algorithm \ref{algo:spatial}.}

\RevOne{For each experiment, a different trajectory of the target \eqref{eq:target} was generated in the interval $t\in[0,T]$ with $T=100$, as well aas different initial conditions for the robots in \eqref{eq:agents}. As performance indicators, use the value of the estimation error $\mfs{RMS}_\mfs{avg}\{\|\hat{\mf{p}}_i(t)-\mf{p}(t)\|\}_{i=1}^\mfs{N}$ and the tracking error $\mfs{RMS}_\mfs{avg}\{\|\bar{\mf{p}}
_{\mfs{G}}(t)+\mf{d}_i-\mf{p}_i(t)\|\}_{i=1}^\mfs{N}$ where recall that $\mf{p}_i(t)$ is the position of the $i$-th robot. In addition, we measure the control effort by means of $\mfs{RMS}_{\mfs{avg}}\{\|\mf{u}_i(t)\|\}_{i=1}^\mfs{N}$ and 
$$
\mfs{PEAK}\{\mf{u}_i(t)\}_{i=1}^\mfs{N}:=\max_{i\in\{1,\dots,\mfs{N}\}}\sup_{t\in[0,T]}\|\mf{u}_i(t)\|
$$}

\RevOne{The results of these simulations are summarized in Table \ref{tab:ablation}. The performance indicators in the rows of Table \ref{tab:ablation} were averaged over all 100 experiments for each column of the table. It can be observed that the DOE performs the best in terms of the estimation error among the configurations which do not use fusion. However, the SOE in conjunction with the fusion method is able to outperform the DOE in all cases, being the best configuration with $\alpha=0.1$ due to the tight consistency of the estimations ensured by Theorem \ref{th:smooth}. The disadvantages of the DOE are more evident when looking at the tracking error, where it performs the worst up to 1 order of magnitude. The reason is that the the discontinuities in the DOE cause the persistent transients illustrated in Figure \ref{fig:network} whereas the SOE does not have this problem in any configuration. }

\RevOne{On the other hand, the bad tracking performance of the DOE is balanced by the resulting small control effort both in RMS and peak values, outperforming the SOE in all cases. The reason is that the the DOE ignores the values of the derivatives of the estimation at the discontinuities since they are undefined for those instants. However, the SOE is able to compute these derivatives for all time. Due to the fact that as $\alpha\to 0$ one recovers a discontinuous behaviour, the derivatives can grow unbounded at the sampling instants as $\alpha$ is decreased. Since these derivatives are used explicitly in the control law \eqref{eq:control}, hence the control effort is impacted by the value of $\alpha$ in a similar fashion. Moreover, an equivalent behaviour happens when $\alpha$ increases, requiring a compromise for this parameter in terms of control effort. For example, the value of $\alpha=1$ obtains good results for the control effort when compared to $\alpha=0.1,10$. }

\begin{table}
\begin{center}
\scalebox{0.86}{
\begin{tabular}{lccccccc}
\toprule
 \multicolumn{1}{c}{Estimator} &  DOE & \multicolumn{2}{c}{$\begin{array}{c}
 \text{SOE}\\
 \alpha=0.1
 \end{array}$} & \multicolumn{2}{c}{$\begin{array}{c}
 \text{SOE}\\
 \alpha=1
 \end{array}$} & \multicolumn{2}{c}{$\begin{array}{c}
 \text{SOE}\\
 \alpha=10
 \end{array}$}\\
 \multicolumn{1}{c}{Fusion}&\multicolumn{1}{c}{\xmark}&\multicolumn{1}{c}{\xmark}&\multicolumn{1}{c}{\cmark}&\multicolumn{1}{c}{\xmark}&\multicolumn{1}{c}{\cmark}&\multicolumn{1}{c}{\xmark}&\multicolumn{1}{c}{\cmark}\\
\midrule
$\mfs{RMS}_\mfs{avg}\{\|\hat{\mf{p}}_i(t)-\mf{p}(t)\|\}_{i=1}^\mfs{N}$ & 0.74 & 0.79 & \textbf{0.19} &0.85 &0.23 & 0.92 & 0.34\\
$\mfs{RMS}_\mfs{avg}\{\|\bar{\mf{p}}
_{\mfs{G}}(t)+\mf{d}_i-\mf{p}_i(t)\|\}_{i=1}^\mfs{N}$ & 56.4 & 2.84 & \textbf{2.81} & 2.82 &2.84 & 2.83& 2.84 \\
$\mfs{RMS}_{\mfs{avg}}\{\|\mf{u}_i(t)\|\}_{i=1}^\mfs{N}$ &\textbf{6.23}&18.43 &22.74 &10.75 &9.24 & 15.22 & 16.81 \\
$\mfs{PEAK}\{\mf{u}_i(t)\}_{i=1}^\mfs{N}$ &\textbf{8.65 }& 70.83 & 75.98 &30.21 &34.45 & 65.01 & 63.78  \\
\bottomrule
\end{tabular}
}
\caption{\RevOne{Ablation and parameter analysis for the proposal. The DOE is compared with with our proposal using different configurations. Moreover, different performance indicators are depicted, where the best value for each row is marked in bold font. In particular the first two rows represent the estimation and tracking RMS errors measured in $\mfs{[m]}$. The last two rows depict the control effort in terms of RMS and peak values, both measured in $[\mfs{m/s}^2]$.}}\label{tab:ablation}
\end{center}
\end{table}

\section{Conclusion}
A combination of a smooth-output estimator and an estimation fusion stage was proposed for distributed target estimation. It was shown that, in contrast to the non-smooth optimal alternative, the formation control is able to achieve asymptotic convergence to the formation goal. This allows the control design to be decoupled from the perception-latency decisions. The advantages of the proposal where discussed through simulation examples, when compared to a non-smooth optimal alternative. \RevOne{The proposal is yet to be validated in real-world platforms, which imposes an interesting but highly non-trivial challenge to be explored in a future work.}

\appendix
\section{Auxiliary matrices}
\label{ap:aux_mat}
Given $m,n\in\mathbb{N}$, let
$$
\mf{A}_0 = \left[\begin{array}{ccccc}
\mf{0}_{m} & \eyem_m \\
0 & \mf{0}_m^\top
\end{array}\right]
$$
$
\mf{B}_0 =[
\mf{0}_m , 1
]^\top
$,
$
\mf{C}_0 = [ 1,  \mf{0}_m^\top
]
$, $\mf{I}_m$ is the identity matrix of $\mathbb{R}^{m\times m}$ and $\mf{0}_m$ is the zero vector in $\mathbb{R}^m$. Furthermore, $\mf{A}=\mf{A}_0\otimes \eyem_n, \mf{B}=\mf{B}_0\otimes \eyem_n, \mf{C} = \mf{C}_0\otimes \eyem_n$ where $\otimes$ is the Kronecker product. Moreover, given $\gamma_0,\dots,\gamma_m\in\mathbb{R}$, let $
\mf{\Gamma} = \mf{A}_0-\diag{\gamma_0,\dots,\gamma_m}
$,
$
\mf{G}= [
(\mf{C}_0)^\top,
(\mf{C}_0\mf{\Gamma})^\top,
\cdots,
(\mf{C}_0\mf{\Gamma}^{m})^\top
]^\top
$.

\section{Example of a transition function}
\label{ap:mod}

\begin{prop}
Let $\alpha>0$ and 
$\upeta(\tau;\alpha) = \dfrac{\Phi(\tau)}{\Phi(\tau) + \Phi(\alpha(1-\tau))}$
with $\Phi(\tau) = \tau^{m+1}$. Then, $\upeta(\tau;\alpha)$ satisfy Definition \ref{def:mod}.
\end{prop}
\begin{pf}
Item \ref{eta:continous} follows from smoothness of $\Phi(\tau)$ leading to smoothness of $\upeta(\tau,\alpha)$ for any $\tau\in[0,1]$.  Note $\Phi^{(\mu)}(0)=0, \forall \mu\in\{1,\dots,m\}$ and hence $\upeta^{(\mu)}(\tau,\alpha)=0$ as well from the product rule. Moreover, $\upeta(\tau;\alpha)=1-\upeta(\alpha(1-\tau);\alpha^{-1})$ so that $\upeta^{(\mu)}(1;\alpha)=-\upeta^{(\mu)}(0;\alpha)=0, \forall \mu\in\{1,\dots,m\}$ showing item \ref{eta:derivatives}. Item \ref{eta:ends} of Definition \ref{def:mod} is complied by evaluating $\upeta(\tau;\alpha)=0, \upeta(\tau;\alpha)=1$. Finally, note that for fixed $\tau\in(0,1)$, it follows that $\lim_{\alpha\to 0}\Phi(\alpha(1-\tau))=0$ point-wise, implying item \ref{eta:pointwise}.
\end{pf}
\section{The REDCHO protocol}
\label{ap:redcho_prot}
Consider $\mfs{N}$ agents connected in a communication network modeled by a graph $\mathcal{G}$. Each agent $i$ has access to a local signal $\hat{s}_i(t)$ and has $m+1$ internal scalar variables ${v}_{i,0}(t),\dots,{v}_{i,m}(t)$ and outputs ${s}_{i,0}(t), \dots, {s}_{i,m}(t)$ for which ${s}_{i,0}(t)$ is communicated to its neighbors. The REDCHO protocol \cite{redcho} of $m$-th order achieves robust Exact Dynamic Consensus (EDC), i.e. the outputs ${s}_{i,\mu}(t)$ converge to the signal
$
\bar{s}^{(\mu)}(t) = (1/\mfs{N})\sum_{i=1}^\mfs{N}\hat{s}_i^{(\mu)}(t),\quad \mu\in\{0,\dots,m\}
$, regardless of spontaneous connection or disconnection of agents as long as the network topology remains connected. Given positive parameters $\{k_\mu,\gamma_\mu\}_{\mu=0}^m,\theta$, the structure of REDCHO is:

\begin{equation}
\label{eq:redcho}
\begin{array}{ll}
    {s}_{i,\mu}(t) &= \hat{s}_i^{(\mu)}(t) - \sum_{\nu=0}^{m} G_{\mu+1,\nu+1} {v}_{i,\nu}(t)\\
    \dot{{v}}_{i,\mu}(t) &= k_\mu\theta^{\mu+1} \sum_{j=1}^\mfs{N}a_{ij}\lceil {s}_{i,0}(t) - {s}_{j,0}(t) \rfloor^{\frac{m-\mu}{m+1}}+\ {v}_{i,\mu+1}(t) - \gamma_\mu {v}_{i,\mu}(t), \\ & \text{for } 0\leq \mu < m\\
    \dot{{v}}_{i,m}(t) &= k_m\theta^{m+1} \sum_{j=1}^\mfs{N}a_{ij}\sgn{{s}_{i,0}(t) - {s}_{j,0}(t) }{0} - \gamma_m{v}_{i,m}(t)
\end{array}
\end{equation}

where $a_{ij}$ are the components of the adjacency matrix of $\mathcal{G}$, $G_{\mu+1,\nu+1}$ with $\mu,\nu\in\{0,\dots,m\}$ are the components of the matrix $\mf{G}$ defined in \ref{ap:aux_mat}. The REDCHO algorithm requires the following assumption.

\begin{assum}
\label{as:u_signals}
Let 
$$\tilde{s}_i(t)=\left(\bar{s}^{(m+1)}(t) - \hat{s}_i^{(m+1)}(t)\right) + \sum_{\mu=0}^m l_\mu \left(\bar{s}^{(\mu)}(t)-\hat{s}_i^{(\mu)}(t)\right)$$
where $l_0,\dots,l_m$ are the coefficients of the polynomial $(\lambda+\gamma_0)\cdots(\lambda+\gamma_m) = \lambda^{(m+1)} + \sum_{\mu=0}^{m}l_\mu \lambda^{\mu}$.
Thus, $|\tilde{s}_i(t)|\leq L, \forall t\geq 0$ {for fixed $\gamma_0,\dots,\gamma_m$ and known $L>0$}.
\end{assum}
\begin{prop} \cite[Adapted from Theorem 10]{redcho}
\label{prop:redcho}
Let  $\mathcal{G}$ be a connected graph and let Assumption \ref{as:u_signals} hold for given $L$. Then, for fixed $\gamma_0,\dots,\gamma_m>0$, there exists $k_0,\dots,k_m,\theta$ sufficiently big, $T>0$, compact sets $\mathcal{R}_0,\dots,\mathcal{R}_m\subset\mathbb{R}^N$ for the initial conditions $\{{s}_{i,\mu}(0)\}_{\mu=0}^m$ to lie and $m$-times differentiable signal $\tilde{s}(t)$ such that \eqref{eq:redcho}  satisfies for all $\mu\in\{0,\dots,m\}$:
\begin{enumerate}[label={\normalfont\textbf{\roman*})}]
    \item $\tilde{s}^{(\mu)}(t)={s}_{1,\mu}(t)=\dots={s}_{\mfs{N},\mu}(t)$, $\forall t\geq T$.
    \item $\lim_{t\to\infty}|\tilde{s}(t)-\bar{s}(t)|=0$.
\item $\mathcal{R}_\mu$ can be made arbitrarily large by increasing $\theta$.
\end{enumerate}
\end{prop}

\section{Proofs}

\subsection{Proof of Theorem  \ref{th:smooth}.}
\label{ap:smooth}
For item \textbf{i)}, note that $\hat{\mf{y}}(t), \hat{\mf{Q}}(t)$ are $m$-times differentiable for any $t\notin\bm{\tau}$ from the expressions in \eqref{eq:estimate_time} and \eqref{eq:estimate} and item \textbf{i)} of Definition \ref{def:mod}. For $t\in\bm{\tau}$, take an arbitrary $\tau_k\in\bm{\tau}$ and compute the $\mu$-th derivative of $\hat{\mf{y}}(t)$ with $\mu\in\{0,\dots,m\}$ as $t\to\tau_k^+$ as:
$$
\begin{aligned}
\lim_{t\to\tau_k^+}\hat{\mf{y}}^{(\mu)}(t) =&\lim_{t\to\tau_k^+} \sum_{\nu=0}^\mu\binom{\mu}{\nu} \uplambda_1^{(\nu)}(t|k)(\hat{\mf{y}}^*)^{(\mu-\nu)}(t|k-1) 
\\&+\binom{\mu}{\nu}\uplambda_2^{(\nu)}(t|k)(\hat{\mf{y}}^*)^{(\mu-\nu)}(t|k) 
= (\hat{\mf{y}}^*)^{(\mu)}(\tau_k|k-1)
\end{aligned}
$$
where the $\mu$-th derivative product rule was used as well as by noting that $\binom{\mu}{0}=1$ and
$$
\begin{aligned}
&\lim_{t\to\tau_k^+}\uplambda_1(t|k) = 1-\upeta(0;\alpha) = 1 \\
&\lim_{t\to\tau_k^+}\uplambda_2(t|k) = \upeta(0;\alpha) = 0 \\
&\lim_{t\to\tau_k^+}\uplambda_1^{(\nu)}(t|k) = -\upeta^{(\nu)}(0;\alpha) = 0 \\
&\lim_{t\to\tau_k^+}\uplambda_2^{(\nu)}(t|k) = \upeta^{(\nu)}(0;\alpha) = 0 \\
\end{aligned}
$$
for any $\nu\in\{1,\dots,m\}$ by items \textbf{ii)} and \textbf{iii)} of Definition \ref{def:mod}. Now, for $t\to\tau_k^-$ consider $ \hat{\mf{y}}(t)=  \uplambda_1(t|k-1)\hat{\mf{y}}^*(t|k-2) + \uplambda_2(t|k-1)\hat{\mf{y}}^*(t|k-1)$ for $t\in[\tau_{k-1},\tau_k)$ from \eqref{eq:estimate}. Similarly as before,
\begin{equation}
\begin{aligned}
\lim_{t\to\tau_k^-}\hat{\mf{y}}^{(\mu)}(t) &=
\lim_{t\to\tau_k^-} \sum_{\nu=0}^\mu\binom{\mu}{\nu} \uplambda_1^{(\nu)}(t|k-1)(\hat{\mf{y}}^*)^{(\mu-\nu)}(t|k-2) \\&+\binom{\mu}{\nu}\uplambda_2^{(\nu)}(t|k-1)(\hat{\mf{y}}^*)^{(\mu-\nu)}(t|k-1) = (\hat{\mf{y}}^*)^{(\mu)}(\tau_k|k-1)
\end{aligned}
\end{equation}
where the following identities were used:
$$
\begin{aligned}
&\lim_{t\to\tau_k^-}\uplambda_1(t|k-1) = 1-\upeta((\tau_k-\tau_{k-1})/\Delta_{k-1};\alpha) = 0 \\
&\lim_{t\to\tau_k^-}\uplambda_2(t|k-1) = \upeta(1;\alpha) = 1 \\
&\lim_{t\to\tau_k^-}\uplambda_1^{(\nu)}(t|k-1) = -\upeta^{(\nu)}(1;\alpha) = 0 \\
&\lim_{t\to\tau_k^-}\uplambda_2^{(\nu)}(t|k-1) = \upeta^{(\nu)}(1;\alpha) = 0 \\
\end{aligned}
$$
due to $(\tau_k-\tau_{k-1}) = \Delta_{k-1}$ and Definition \ref{def:mod}. Hence, the two-sided limit $\lim_{t\to\tau_k}\hat{\mf{y}}^{(\mu)}(t)= (\hat{\mf{y}}^*)^{(\mu)}(\tau_k|k-1)$ exists for arbitrary $\tau_k\in\bm{\tau}$. The proof is the same for $\hat{\mf{Q}}(t)$.
For item \textbf{ii)} note that $\hat{\mf{x}}(t|k)$ is unbiased since it comes from the Kalman filter structure of  \eqref{eq:kalman} and \eqref{eq:estimate_time}. Hence, $\hat{\mf{x}}(t)$ is unbiased by linearity of $\mathbb{E}\{\bullet\}$ and by the definition of $\hat{\mf{x}}(t)$ from \eqref{eq:estimate} as a convex combination of unbiased estimates since $\uplambda_1(t|k)+\uplambda_2(t|k) = 1$.
For item \textbf{iii)} note that $\hat{\mf{P}}^*(t|k),\hat{\mf{P}}^*(t|k-1)$ for $t\in[\tau_k,\tau_{k+1})$ are covariances for $\hat{\mf{x}}^*(t|k),\hat{\mf{x}}^*(t|k-1)$ but that $\cov\{\mf{x}(t)-\hat{\mf{x}}^*(t|k),\mf{x}(t)-\hat{\mf{x}}^*(t|k-1)\}\neq 0$ since both estimations are correlated by their dependence of the noise process $\{\mf{u}(t') | t'\leq t\}$ in \eqref{eq:target}. However, note that the expression of $\hat{\mf{Q}}(t)$ is a convex combination of $\hat{\mf{Q}}^*(t|k),\hat{\mf{Q}}^*(t|k-1)$. Hence, the covariance-intersection principle from \cite[Section 2.1]{cov_intersection} ensures consistency of $\hat{\mf{P}}(t)$ i.e., $\hat{\mf{P}}^*(t)\preceq\hat{\mf{P}}(t)$. Finally, note that by letting $\alpha\to 0$ we have $\uplambda_1(t|k)=0, \uplambda_1(t|k)=1$ for all $t\in(\tau_k,\tau_{k+1})$ using item \textbf{iv)} from Definition \ref{def:mod}. Hence, $\lim_{\alpha\to 0}\hat{\mf{P}}(t) = \hat{\mf{P}}^*(t|k) = \hat{\mf{P}}^*(t), \forall t\notin\bm{\tau}$.

\subsection{Proof of Lemma \ref{le:derivatives}}
\label{ap:derivatives}
For item \textbf{i)}, take the $\mu$-th derivative of the identity $\mf{I}_m=\bar{\mf{Q}}_{\mfs{G}}(t)\bar{\mf{P}}_{\mfs{G}}(t)$ yielding:
$$
\begin{aligned}
0&=\sum_{\nu=0}^\mu\binom{\mu}{\nu} \bar{\mf{Q}}_{\mfs{G}}^{(\mu-\nu)}(t)\bar{\mf{P}}_{\mfs{G}}^{(\nu)}(t)=\bar{\mf{Q}}_{\mfs{G}}(t)\bar{\mf{P}}_{\mfs{G}}^{(\mu)}(t)+ \sum_{\nu=0}^{\mu-1}\binom{\mu}{\nu} \bar{\mf{Q}}_{\mfs{G}}^{(\mu-\nu)}(t)\bar{\mf{P}}_{\mfs{G}}^{(\nu)}(t)
\end{aligned}
$$
where $\binom{\mu}{\mu}=1$ was used and from which the expression in item \textbf{i)} is obtained by solving for $\bar{\mf{P}}_{\mfs{G}}^{(\mu)}(t)$. Finally, item \textbf{ii)} is obtained by applying the $\mu$-th derivative to the definition $\bar{\mf{p}}_{\mfs{G}}(t)=\mf{C}\bar{\mf{P}}_{\mfs{G}}(t)\bar{\mf{y}}_{\mfs{G}}$, completing the proof.
\subsection{Proof of Theorem \ref{th:spatial}}
\label{ap:spatial}

First, note that each component of the equations in \eqref{eq:redcho_vector} and \eqref{eq:redcho_matrix} is a REDCHO block of the form \eqref{eq:redcho}. Each individual REDCHO block in \eqref{eq:redcho_vector} (resp. \eqref{eq:redcho_matrix}) has scalar input $\hat{s}_i(t)$ given by one component of $\hat{\mf{y}}_{i}(t)$ (resp.  $\hat{\mf{Q}}_{i}(t)$), internal scalar variables $\{v_{i,\mu}(t)\}_{\mu=0}^m$ given by one component of each $\{\mf{v}_{i,\mu}(t)\}_{\mu=0}^m$ (resp. $\{\mf{V}_{i,\mu}(t)\}_{\mu=0}^m$) and outputs $\{s_{i,\mu}(t)\}_{\mu=0}^m$ given by one component of each $\{\mf{y}_{i,\mu}(t)\}_{\mu=0}^m$ (resp. $\{\mf{Q}_{i,\mu}(t)\}_{\mu=0}^m$). Hence Assumption together with item \textbf{i)} of Proposition \ref{prop:redcho} imply that there exists $T>0$ such that
$$
\begin{aligned}
\tilde{\mf{y}}^{(\mu)}(t)&=\mf{y}_{1,\mu}(t)=\cdots=\mf{y}_{\mfs{N},\mu}(t) \\
\tilde{\mf{Q}}^{(\mu)}(t)&=\mf{Q}_{1,\mu}(t)=\cdots=\mf{Q}_{\mfs{N},\mu}(t)
\end{aligned}
$$
for all $\mu\in\{0,\dots,m\}$ for some consensus signals $\tilde{\mf{y}}(t)\in\mathbb{R}^{nm}, \tilde{\mf{Q}}(t)\in\mathbb{R}^{nm\times nm}$. Moreover, define $\tilde{\mf{P}}(t)=\tilde{\mf{Q}}(t)^{-1}$ Hence, for all $t\geq T$ and $i\in\{1,\dots,N\}$ the expressions in Algorithm \ref{algo:spatial} reads:
\begin{equation}
\label{eq:algo_convergence}
    \begin{aligned}
    \mf{p}_{i,0}(t)&\equiv \mf{C}\tilde{\mf{P}}(t)\tilde{\mf{y}}_0(t)\\
    \mf{P}_{i,\mu}(t)&\equiv -\tilde{\mf{P}}(t)\sum_{\nu=0}^{\mu-1}\binom{\mu}{\nu}\tilde{\mf{Q}}^{(\mu - \nu)}(t)\mf{P}_{i,\nu}(t)\\
    \mf{p}_{i,\mu}(t)&\equiv \mf{C}\sum_{\nu=0}^{\mu}\binom{\mu}{\nu}\mf{P}_{i,\nu}(t)\tilde{\mf{y}}^{(\mu-\nu)}(t)
    \end{aligned}
\end{equation}
Now, item \textbf{ii)} of Proposition \ref{prop:redcho} imply that $\tilde{\mf{y}}(t),\tilde{\mf{Q}}(t)$ converge to $\bar{\mf{y}}_{\mfs{G}}(t),\bar{\mf{Q}}_{\mfs{G}}(t)$ as $t\to\infty$. Hence, the expressions in \eqref{eq:algo_convergence} converge to $\bar{\mf{p}}_{\mfs{G}}(t), \bar{\mf{P}}_{\mfs{G}}^{(\mu)}(t), \bar{\mf{p}}_{\mfs{G}}^{(\mu)}(t)$ respectively due to Lemma \ref{le:derivatives}, completing the proof.

\bibliographystyle{elsarticle-num}

\begin{thebibliography}{10}
\expandafter\ifx\csname url\endcsname\relax
  \def\url#1{\texttt{#1}}\fi
\expandafter\ifx\csname urlprefix\endcsname\relax\def\urlprefix{URL }\fi
\expandafter\ifx\csname href\endcsname\relax
  \def\href#1#2{#2} \def\path#1{#1}\fi

\bibitem{escort}
Y.~Lan, Z.~Lin, M.~Cao, G.~Yan, A distributed reconfigurable control law for
  escorting and patrolling missions using teams of unicycles, in: 49th IEEE
  Conference on Decision and Control (CDC), 2010, pp. 5456--5461.

\bibitem{fusion_trajectory}
F.~Castanedo, J.~García, M.~A. Patricio, J.~M. Molina, Data fusion to improve
  trajectory tracking in a cooperative surveillance multi-agent architecture,
  Information Fusion 11~(3) (2010) 243--255, agent-Based Information Fusion.

\bibitem{film}
L.-E. Caraballo, A.~Montes-Romero, J.-M. Diaz-Banez, J.~Capitan,
  A.~Torres-Gonzalez, A.~Ollero, Autonomous planning for multiple aerial
  cinematographers, in: IEEE/RSJ International Conference on Intelligent Robots
  and Systems (IROS), 2020, pp. 1509--1515.

\bibitem{transport}
N.~Michael, J.~Fink, V.~Kumar, Cooperative manipulation and transportation with
  aerial robots, Auton. Robots 30~(1) (2011) 73–86.

\bibitem{redmon2016}
J.~Redmon, S.~K. Divvala, R.~B. Girshick, A.~Farhadi, You only look once:
  Unified, real-time object detection, IEEE Conference on Computer Vision and
  Pattern Recognition (CVPR) (2016) 779--788.

\bibitem{beery2020}
S.~Beery, G.~Wu, V.~Rathod, R.~Votel, J.~Huang, Context r-cnn: Long term
  temporal context for per-camera object detection, IEEE/CVF Conference on
  Computer Vision and Pattern Recognition (CVPR) (2020) 13072--13082.

\bibitem{ren2015}
S.~Ren, K.~He, R.~Girshick, J.~Sun, Faster r-cnn: Towards real-time object
  detection with region proposal networks, IEEE Transactions on Pattern
  Analysis and Machine Intelligence 39 (06 2015).

\bibitem{Dai2016}
J.~Dai, Y.~Li, K.~He, J.~Sun, R-fcn: Object detection via region-based fully
  convolutional networks, in: Proceedings of the 30th International Conference
  on Neural Information Processing Systems, NIPS'16, Curran Associates Inc.,
  Red Hook, NY, USA, 2016, p. 379–387.

\bibitem{cyber_rnn}
L.~Wang, L.~Zhang, Z.~Yi, Trajectory predictor by using recurrent neural
  networks in visual tracking, IEEE Transactions on Cybernetics 47~(10) (2017)
  3172--3183.

\bibitem{aldana2020}
R.~Aldana-López, R.~Aragüés, C.~Sagüés, Attention vs. precision: latency
  scheduling for uncertainty resilient control systems, in: IEEE Conference on
  Decision and Control (CDC), 2020, pp. 5697--5702.

\bibitem{huang2017}
J.~Huang, V.~Rathod, C.~Sun, M.~Zhu, A.~Korattikara, A.~Fathi, I.~Fischer,
  Z.~Wojna, Y.~Song, S.~Guadarrama, K.~Murphy, Speed/accuracy trade-offs for
  modern convolutional object detectors, in: 2017 {IEEE} Conference on Computer
  Vision and Pattern Recognition, {CVPR} 2017, Honolulu, HI, USA, July 21-26,
  2017, 2017, pp. 3296--3297.

\bibitem{luo2019}
H.~Luo, W.~Xie, X.~Wang, W.~Zeng, Detect or track: Towards cost-effective video
  object detection/tracking, Proceedings of the AAAI Conference on Artificial
  Intelligence 33 (2019) 8803--8810.

\bibitem{guan2018}
M.~Guan, C.~Wen, M.~Shan, C.-L. Ng, Y.~Zou, Real-time event-triggered object
  tracking in the presence of model drift and occlusion, IEEE Transactions on
  Industrial Electronics PP (05 2018).

\bibitem{scheduling}
S.~Yao, Y.~Hao, Y.~Zhao, H.~Shao, D.~Liu, S.~Liu, T.~Wang, J.~Li,
  T.~Abdelzaher, Scheduling real-time deep learning services as imprecise
  computations, in: IEEE 26th International Conference on Embedded and
  Real-Time Computing Systems and Applications (RTCSA), 2020, pp. 1--10.

\bibitem{hu2019}
H.~Hu, D.~Dey, M.~Hebert, J.~Bagnell, Learning anytime predictions in neural
  networks via adaptive loss balancing, Proceedings of the AAAI Conference on
  Artificial Intelligence 33 (2019) 3812--3821.

\bibitem{fusion_filtering1}
H.~Jin, S.~Sun, Distributed filtering for multi-sensor systems with missing
  data, Information Fusion (2022).

\bibitem{fusion_filtering2}
S.-L. Sun, Multi-sensor optimal fusion fixed-interval kalman smoothers,
  Information Fusion 9~(2) (2008) 293--299.

\bibitem{static_consensus}
A.~T. Kamal, J.~A. Farrell, A.~K. Roy-Chowdhury, Information weighted consensus
  filters and their application in distributed camera networks, IEEE
  Transactions on Automatic Control 58~(12) (2013) 3112--3125.

\bibitem{olfati1}
R.~Olfati-Saber, Distributed kalman filter with embedded consensus filters, in:
  IEEE Conference on Decision and Control, 2005, pp. 8179--8184.

\bibitem{olfati2}
R.~Olfati-Saber, Distributed kalman filtering for sensor networks, in: IEEE
  Conference on Decision and Control, 2007, pp. 5492--5498.

\bibitem{Solmaz2017}
S.~S. {Kia}, B.~{Van Scoy}, J.~{Cortes}, R.~A. {Freeman}, K.~M. {Lynch},
  S.~{Martinez}, Tutorial on dynamic average consensus: The problem, its
  applications, and the algorithms, IEEE Control Systems Magazine 39~(3) (2019)
  40--72.

\bibitem{edcho}
R.~Aldana-López, R.~Aragüés, C.~Sagüés, {EDCHO}: High order exact dynamic
  consensus, Automatica 131 (2021) 109750.

\bibitem{CI1}
S.~Wang, W.~Ren, On the convergence conditions of distributed dynamic state
  estimation using sensor networks: A unified framework, IEEE Transactions on
  Control Systems Technology 26~(4) (2018) 1300--1316.

\bibitem{CI2}
X.~He, W.~Xue, H.~Fang, Consistent distributed state estimation with global
  observability over sensor network, Automatica 92 (2018) 162--172.

\bibitem{CI3}
E.~Sebastian, E.~Montijano, C.~Sagues, All-in-one: Certifiable optimal
  distributed kalman filter under unknown correlations, in: IEEE Conference on
  Decision and Control (CDC), 2021, pp. 6578--6583.

\bibitem{redcho}
R.~Aldana-López, R.~Aragüés, C.~Sagüés, {REDCHO}: Robust exact dynamic
  consensus of high order, Automatica 141 (2022) 110320.

\bibitem{astrom}
K.~{\AA}str{\"o}m, Introduction to Stochastic Control Theory, Mathematics in
  science and engineering, Academic Press, 1970.

\bibitem{codesign}
Y.~V. Pant, H.~Abbas, K.~Mohta, R.~A. Quaye, T.~X. Nghiem, J.~Devietti,
  R.~Mangharam, Anytime computation and control for autonomous systems, IEEE
  Transactions on Control Systems Technology 29~(2) (2021) 768--779.

\bibitem{Soderstrom}
T.~Soderstrom, Discrete-Time Stochastic Systems: Estimation and Control, 2nd
  Edition, Springer-Verlag, Berlin, Heidelberg, 2002.

\bibitem{song2019}
Y.-M. Song, K.~Yoon, Y.~Young~Chul, K.-C. Yow, M.~Jeon, Online multi-object
  tracking with gmphd filter and occlusion group management, IEEE Access PP
  (2019) 1--1.

\bibitem{cov_intersection}
W.~Niehsen, Information fusion based on fast covariance intersection filtering,
  in: International Conference on Information Fusion., Vol.~2, 2002, pp.
  901--904 vol.2.

\bibitem{kalman_ros}
R.~Aragues, C.~Sagues, Y.~Mezouar, Feature-based map merging with dynamic
  consensus on information increments, in: IEEE International Conference on
  Robotics and Automation, 2013, pp. 2725--2730.

\end{thebibliography}

\end{document}